\documentclass[aps,physrev,twocolumn,superscriptaddress]{revtex4-2}

\usepackage{graphicx}
\usepackage{dcolumn}
\usepackage{bm}
\usepackage{color}
\usepackage[normalem]{ulem}
\usepackage{soul}
\usepackage{mathtools}
\usepackage{amsmath}
\usepackage{amssymb}

\def\CITE#1{{\color{red}[CITE]}}
\newcommand{\DD}{\triad {\tilde{D}}kpq}
\newcommand{\vv}{\mathbf{v}}
\newcommand{\kk}{\mathbf{k}}
\newcommand{\pp}{\mathbf{p}}
\newcommand{\qq}{\mathbf{q}}
\newcommand{\triad}[4]{#1^{\mathbf{#2}}_{\mathbf{#3}\mathbf{#4}}}
\newcommand{\tphase}{\triad \theta kpq}
\newcommand{\CC}{\triad Ckpq}
\newcommand{\KK}{\triad {\mathcal{K}}kpq}
\newcommand{\XX}{\triad \xi kpq}
\newcommand{\XXs}{\triad {\tilde{\xi}} kpq}

\usepackage{hyperref}

\begin{document}


\title{Triad phase dynamics determine cascade direction in two-dimensional turbulence}

\author{Santiago J. Benavides}
\email[]{Santiago.Benavides@ed.ac.uk}
\affiliation{School of Mathematics and Maxwell Institute for Mathematical Sciences, University of Edinburgh, Edinburgh, UK}
\affiliation{School of Aeronautics and Space Engineering, Universidad Polit\'{e}cnica de Madrid, Madrid, Spain}

\author{Miguel D. Bustamante}
\email[]{miguel.bustamante@ucd.ie}
\affiliation{School of Mathematics and Statistics, University College Dublin, Belfield, Dublin 4, Ireland}

\date{\today}

\begin{abstract}
Despite their importance in turbulence theory, a unifying and predictive rule determining the direction of the cascades of conserved quantities is lacking. 
In this work, we show that the direction of the cascades in two-dimensional turbulence is encoded in the complex phases of the Fourier transform of the velocity field. We develop a closure for the dynamics of a triad phase, the sum of the phases of three modes forming a triad, based on the observation that neighboring triad phases are weakly correlated. 
The resulting stochastic model can be solved analytically to find the triad phase probability distribution function (PDF). 
We validate our model's assumptions and predictions using an ensemble of two-dimensional turbulence simulations. 
From the triad phase PDF we develop a novel closure of the energy equation, and prove that the cascade directions are determined by our model without adjustable parameters and given only the energy spectrum. 
Triad phase dynamics occur in any quadratically nonlinear partial differential equation, making this a promising new direction in the study of strongly out-of-equilibrium systems.
\end{abstract}


\maketitle

\emph{Introduction---}One of the most celebrated paradigms in turbulence centers on the concept of the energy cascade in homogeneous and isotropic turbulence \cite{FrischBook,AlexakisReview}.
It posits the existence of a so-called `inertial range', within which a constant energy flux brings energy from the forcing scale, where it is injected, to the dissipation range, where it is removed from the system. 
In incompressible, homogeneous and isotropic three-dimensional (3D) turbulent flows, energy flows towards scales smaller than the forcing scale via the `forward' cascade \cite{FrischBook,AlexakisReview,LesieurBook}. Two-dimensional (2D) incompressible turbulent flows behave strikingly different, forming instead an `inverse' cascade where energy flows to larger scales \cite{Kraichnan1967,BoffettaReview,AlexakisReview}. 
While the forward cascade of 3D turbulence and the inverse cascade of 2D turbulence are well established based on experimental and numerical evidence,  the presence and direction of cascades in anisotropic turbulence has yet to be understood \cite{vanKan2024,AlexakisReview,MarstonReview,Alexakis2023}. 

What determines the direction of the energy cascade is an active area of research spanning decades.
Most approaches, in both configuration \cite{Eyink20062DCL,Chen2006,Xiao2009,Eyink2006MSG, Doan2018,Carbone2020,Johnson2020,Johnson2021,JohnsonReview,Park2025,Ballouz2018,Ballouz2020} and Fourier (spectral) \cite{Fjortoft1953,Kraichnan1967,Kraichnan1973,Kraichnan1975,Waleffe1992,Waleffe1993} space, give arguments based on the conserved quantities of the system (which differ between 2D and 3D turbulence)  or on kinematic considerations. While much has been learned about the constraints on energy cascades, as well as the physical mechanisms that drive them, these approaches are often case-specific and therefore lack a unifying dynamical mechanism responsible for determining the presence and direction of an energy cascade.

The complex phases of the Fourier transform of the velocity field provide one such mechanism.
Consider the velocity field of an incompressible 2D fluid in a periodic domain, which we write in terms of the streamfunction $\psi(\mathbf{x},t)$: $\vv(\mathbf{x},t)= (\partial_y\psi,-\partial_x \psi)$, where $\mathbf{x}=(x,y)$. Taking the Fourier transform of the streamfunction results in a complex-valued function of time and the wavevector, or mode, $\kk = (k_x,k_y)$, which we express in amplitude-phase representation $\widehat{\psi}(\kk,t) = \rho_\kk\exp{(i \phi_\kk)}$ using the shorthand $\rho_{\kk}:= \rho(\kk,t)$, and $\phi_\kk :=\phi(\kk,t)$.
Inviscid 2D flows have two quadratic conserved quantities: the energy $E = \int \vv^2 d\mathbf{x}$ and the enstrophy $\Omega=\int \omega_z^2 d\mathbf{x}$, where $\omega_z = \partial_x v_y - \partial_y v_x = -\nabla^2 \psi$ is the vorticity. The energy forms an inverse cascade whereas the enstrophy forms a forward cascade. 
The energy flux at a scale $2 \pi/ K$, $\Pi_E(K)$, is the rate of energy
loss from modes with $k < K$ due to nonlinear interactions, where $k:=|\kk|$. A positive $\Pi_E(K)$ represents a forward flux at $K$. The flux can be written as:
\begin{equation}
   \Pi_E(K) := \frac{1}{2}\sum_{\kk, \, k<K}\sum_{\pp,\qq} \delta(\kk+\pp+\qq) \triad tkpq , \label{eq:flux_def}  
\end{equation}
where $\triad tkpq$ is the transfer function,
\begin{equation}
    \triad t k p q := (q^2-p^2) (\qq\times\pp) \rho_\kk \rho_\pp \rho_\qq \cos{\left(\tphase\right)} \label{eq:transf_def},
\end{equation} 
and $\tphase$ is the \textit{triad phase},
\begin{equation}
    \tphase := \phi_\kk+\phi_\pp+\phi_\qq. \label{eq:theta_def}
\end{equation}
The triad phase is a natural variable choice, as it is the only form in which the phases appear in the evolution equations of $\rho_{\kk}$ and $\phi_{\kk}$. 
Importantly, the triad phase plays a central role in the energy flux.
Depending on the statistical behavior of $\tphase$, the flux can be zero (e.g., if $\tphase$ is random-uniform) or change sign, thereby determining its strength and direction.
The phases of the modes in a triad must therefore align across scales to facilitate a non-zero energy flux. Finding a closure for the statistics of the triad phase (and therefore the transfer function) in terms of amplitudes $\rho_\kk$ is one of the many challenges in turbulence theory \cite{Orszag1970}.

The important role of the triad phase in determining energy flux has already been recognized and quantified in previous studies of one-dimensional (1D) models of turbulence, such as Burgers and shell models. They have shown that phases in a triad do indeed align to sustain an energy flux, and that large energy transfers are also facilitated by a strong increase in phase alignment \cite{Eyink2003,Buzzicotti2016,Moradi2017,Murray2018,Arguedas2022,CarrollThesis,Protas2024,Benavides2026,Manfredini2026}. Shell models in particular, where both forward and inverse cascades can be produced, show that a change in the probability distribution function (PDF) of $\tphase$ is responsible for the change in cascade direction \cite{Benavides2026,Manfredini2026}.
More recently, work on 2D and 3D turbulence have studied the role of phase dynamics in extreme energy transfers\cite{Kang2021}, dissipation events\cite{Reynolds2016}, and decay\cite{Wang2024}. 
Beyond isotropic homogeneous turbulent flows, the triad phase (called the biphase in these contexts) has also been measured to quantify nonlinear energy transfers \cite{Kim1980,Cui2021}.

Despite having ample evidence of the crucial role that the triad phase plays in determining the strength and direction of the energy cascade, this has not translated to predictions or constraints on the energy cascade in real turbulent systems. This is partly due to the complex nature of the dynamical equation of the triad phase, wherein one triad interacts with its many neighboring triads through nonlinear terms.
In this Letter, we make a crucial, yet sensible simplifying assumption to the dynamics of the triad phases for 2D turbulence, resulting in an analytical approximation for the statistics of the triad phases at steady state, and therefore a closure for the transfer function in terms of mode amplitudes. 
As a result, our triad phase dynamical model determines the energy and enstrophy cascade directions with no fitting parameters, and gives information about the cascades' existence and strength in terms of the average energy spectrum.
Our work provides a direct connection between triad phase dynamics and cascades , and gives a clear path forward for studying cascades in other turbulent systems.

\emph{Triad phase dynamics and model---}We begin by writing down the evolution equation for the triad phase $\tphase$,
\begin{equation}
    \frac{d \tphase}{dt} = \triad C\kk\pp\qq \sin{\left(\tphase\right)} + \sum_{\mathbf{k^\prime} = \{\kk,\pp,\qq\}}\sum^\Delta_{\mathbf{l},\mathbf{r}} \triad {N}{k^\prime}lr \sin{\left(\triad \theta {k^\prime} lr \right)}, \label{eq:triad_phase_evolution}
\end{equation}
which contains a `self-interaction' term with coefficient
\begin{equation}
    \triad C\kk\pp\qq := -\frac{\qq \times \pp}{\rho_\kk \rho_\pp \rho_\qq} \KK, \quad \KK := \sum_{\substack{\circlearrowleft \\ \{\kk,\pp,\qq\}}}\frac{q^2-p^2}{k^2}(\rho_\pp \rho_\qq)^2 \label{eq:Cdef},
\end{equation}
and neighboring triad terms with coefficients $\triad {N}{k^\prime}lr = (\mathbf{r}\times \mathbf{l}) r^2\rho_{\mathbf{r}}\rho_{\mathbf{l}}/((k^\prime)^2\rho_\mathbf{k^\prime})$.
The $\Delta$ in the sum in Eq. \eqref{eq:triad_phase_evolution} represents sums over $\mathbf{l}$ and $\mathbf{r}$ which form a triad with $\mathbf{k^\prime}$.
Given the chaotic and complex nature of the solutions to 
Eq. \eqref{eq:triad_phase_evolution}, our goal will be to model the statistics of triad phases. 
To do so, we introduce our key modeling assumption, following previous work by the authors on shell models of turbulence \cite{Benavides2026}: we will treat the sum of all neighboring triads in Eq. \eqref{eq:triad_phase_evolution} as a single noise variable $\XX$. In doing so, we implicitly neglect any statistical dependence between pairs of triad phases, and invoke the central limit theorem to justify treating $\XX$ as a Gaussian variable. 
While the correlation between triad phases is known to be responsible for intermittency and extreme energy flux events \cite{Eyink2003,Murray2018,CarrollThesis,Manfredini2026}, we will show below that correlations between neighboring triad phases are weak enough that our model does indeed provide a sufficiently accurate description of the steady state statistics of $\tphase$ and its implications on the flux, which are our main goals for this work.

The resulting model equation for the triad phase dynamics is that of a noisy phase oscillator, 
\begin{equation}
    \frac{d \tphase}{dt} = \triad C\kk\pp\qq \sin{\left(\tphase\right)} + \XX. \label{eq:noisyAdler}
\end{equation}
Its deterministic fixed points are $\tphase= \{0,\pm \pi\}$, with the former being stable when $\triad C\kk\pp\qq<0$ and the latter when $\triad C\kk\pp\qq>0$. Assuming $\XX$ is Gaussian white noise with zero mean and variance $\triad Dkpq$, $\overline{\XX(t)\XX(t+s)} = 2 \triad Dkpq \delta(s)$, and ignoring fluctuations of $\CC$\footnote{This is justified under the assumption that the evolution of the amplitudes tends to be slower than that of the phases \cite{Tennekes1975,LesieurBook}.}, we can solve the Fokker-Planck equation for the steady-state PDF of $\tphase$, 
\begin{equation}
    \mathcal{P}\left(\tphase\right) = \frac{1}{2 \pi I_0\left[\frac{\CC}{\triad Dkpq}\right]} \exp{\left(-\frac{\CC}{\triad Dkpq} \cos\left(\tphase\right)\right)}, \label{eq:theta_PDF}
\end{equation}
where $I_n$ is the modified Bessel function of the first kind (order $n$) \cite{ArfkenMathMethodsBook}.
From the PDF we can calculate the mean of $\cos(\tphase)$, which appears in the flux and which we call the \textit{alignment} of the triad, $\langle \cos(\tphase) \rangle = I_1[-\CC/\triad Dkpq]/I_0[\CC/\triad Dkpq]$. The sign of $\langle \cos(\tphase) \rangle$ is entirely determined by $\CC$, since $\triad Dkpq >0$ and the maximum of the PDF occurs at the stable fixed points of the deterministic equation. In particular, we find that $|\CC| \ll \triad Dkpq$, resulting in $\langle \cos(\tphase) \rangle \approx -\CC/\left(2 \triad Dkpq\right)$.

\begin{figure}
    \centering
    \includegraphics[width=\linewidth]{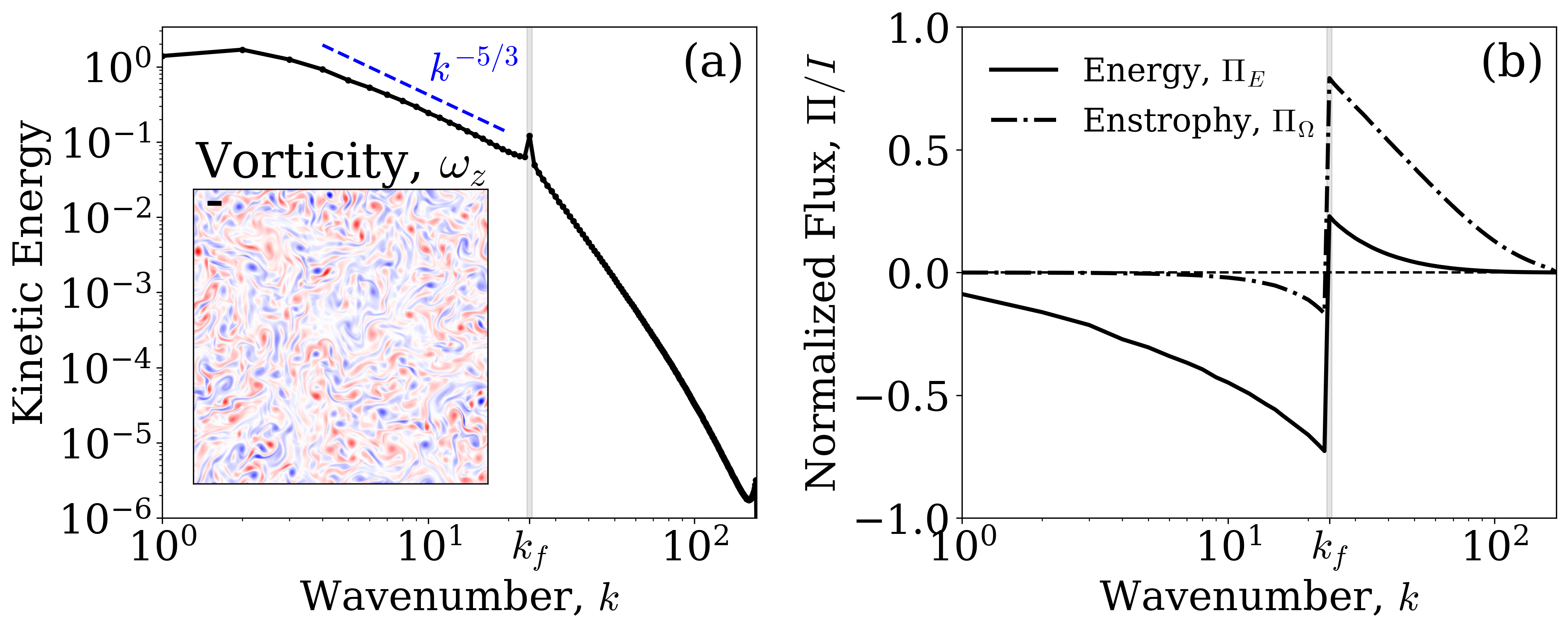}
    \caption{Simulation of two-dimensional turbulence with resolution $512^2$, forced at $k_f = 24$. (a) Energy spectrum, averaged in time and over ensemble members. 
    Inset: a snapshot of the vorticity, $\omega_z$, from an ensemble member. (b) The energy flux, $\Pi_E$ (solid line) and enstrophy flux $\Pi_\Omega$ (dot-dashed line) normalized by their respective injection rates.}
    \label{fig:spec_flux}
\end{figure}
\emph{Testing model assumptions and predictions---}We test the assumptions and predictions of our model for triad phase dynamics using an ensemble of ten pseudo-spectral simulations of 2D turbulence in a periodic domain with side lengths  $2 \pi$, viscous dissipation $\nu \nabla^2 \vv$, and a drag term $-\mu \vv$ acting as a large-scale energy sink (Fig. \ref{fig:spec_flux}) \cite{Benavides2026CodeGPU}. The simulations are forced at an intermediate wavenumber $k_f = 24$, resulting in a drag-based Reynolds number $Re_\mu = 119.1$ (see End Matter for details on the simulations).
Each ensemble member was initialized with a different random initial condition, and statistics were gathered over ensemble members and time once a steady state was reached.
We collected time-series data of $\tphase$ and the supposed `noise' variable \footnote{A tilde is used to differentiate between the true (hypothesised) random variable $\XX$ and the variable $\XXs$ which is measured from the numerical simulations. The latter is supposed to approximate a true noise, but is in fact deterministic.} $\XXs := \sum_{\mathbf{k^\prime} = \{\kk,\pp,\qq\}}\sum^\Delta_{\mathbf{l},\mathbf{r}} \triad {N}{k^\prime}lr \sin{(\triad \theta {k^\prime} lr)}$ for 48 triads, and computed on-the-fly histograms of $\tphase$ for 8657 triads. The triads considered were all randomly chosen, and were taken from both energy and enstrophy cascade ranges, but did not have legs in both. 

\begin{figure}
    \centering
    \includegraphics[width=\linewidth]{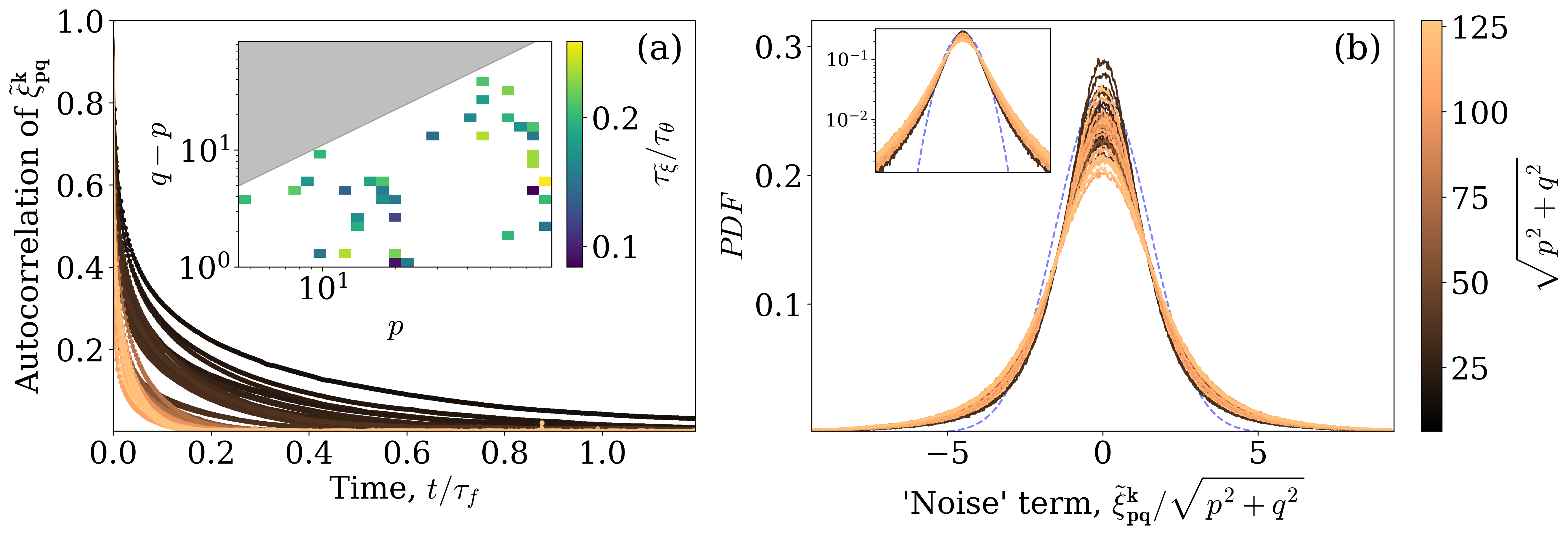}
    \caption{Statistical properties of `noise' variable $\XXs$ for 48 triads. (a) Autocorrelation function shows decorrelation over less than an eddy turnover time $\tau_f$. Inset: ratio of autocorrelation times of the noise variable, $\tau_{\tilde{\xi}}$, and the triad phase, $\tau_\theta$, in scale. (b) PDF of $\XXs$ normalized by $\sqrt{p^2+q^2}$ compared to a Gaussian (dashed blue line). Inset: the same but with log vertical axis.}
    \label{fig:noise}
\end{figure}
To validate the model assumptions, we must verify that the measured `noise' term $\XXs$ can be approximated by a white, Gaussian noise. 
In Figure \ref{fig:noise}(a), we show the temporal autocorrelation of $\XXs$ for all triads, which decay towards zero on a time-scale comparable to the nonlinear eddy turnover time $\tau_f :=(k_f v_{k_f,\mathrm{rms}})^{-1}$, where $v_{k_f,\mathrm{rms}}$ is the square root of the time- and space-mean of the energy based on a velocity spectrally filtered around $k_f -0.5 < k < k_f + 0.5$. The inset shows the ratio of integral time-scales of the noise $\tau_\xi$ and the triad phase $\tau_\theta$, calculated from the integral of the autocorrelation functions. We see that, regardless of the scale of the triad, $\tau_\xi \lesssim 0.25 \tau_\theta$, approximately justifying the white-in-time assumption. 
In Figure \ref{fig:noise}(b), we show the PDF of $\XXs$ for each triad normalized by $\sqrt{p^2+q^2}$, which approximately collapses the PDFs. The PDFs possess heavy tails with power-law exponents of approximately $-2.8$ (not shown). This suggests $\XX$ is better modelled as a L\'{e}vy white noise process, whose PDF is an $\alpha$-stable distribution with $\alpha=1.8$ and zero skewness \cite{ChechkinBook}. 
In the End Matter, we show that the approximate forms for $\mathcal{P}(\tphase)$ in the L\'{e}vy noise case and the white noise case, Eq. \eqref{eq:theta_PDF}, are identical up to first order in $\CC/\triad Dkpq$.
We therefore continue to consider Gaussian white noise in our model for simplicity, whose variance we will measure from the simulation PDF of $\tphase$ and call $\DD$. 

\begin{figure*}
    \centering
    \includegraphics[width=\linewidth]{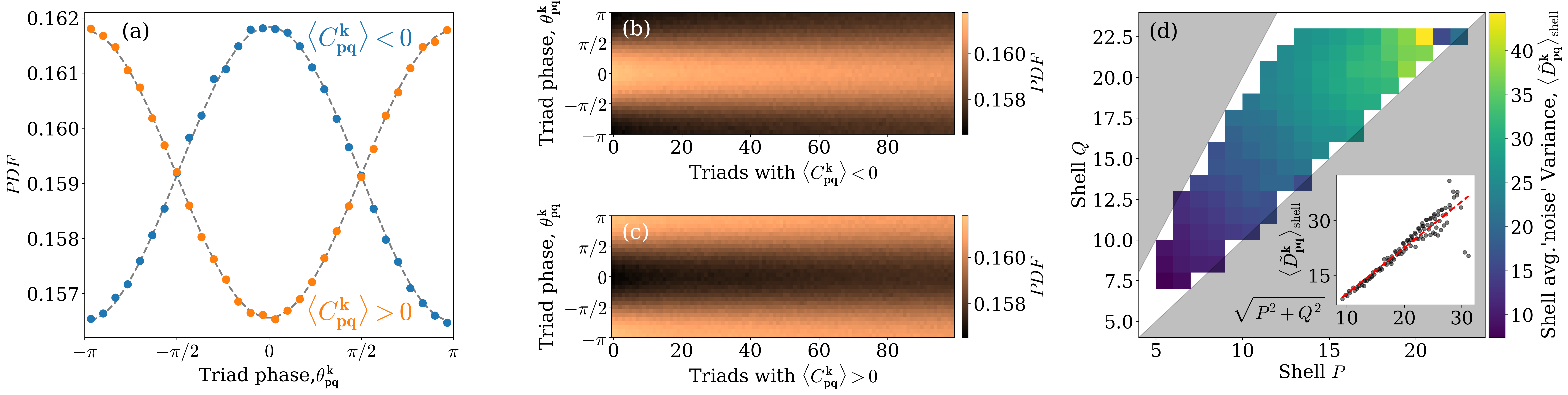}
    \caption{
    Triad phase statistics. (a) Triad phase PDFs for two representative triads, one with $\langle \CC\rangle < 0$ (blue) and the other with $\langle \CC \rangle>0$ (orange). Grey dashed lines show a fit with the theoretical PDF, Eq. \eqref{eq:theta_PDF}. 
    (b) Triad phase PDFs for 100 triads with $\langle \CC \rangle<0$,  shown in descending order of $|\langle \cos(\tphase)\rangle|$. (c) same as (b) but for $\langle \CC \rangle>0$.
    (d) The `noise' variable variance $\DD$ averaged over triads with constant $p=P$ and $q=Q$ (shells). Grey areas denote regions where triads cannot be formed, since we have ordered our triads such that $k<p<q$. Inset: The same quantity shown as a function of $\sqrt{P^2+Q^2}$, showing its linear relation (red dashed line).
    }
    \label{fig:triads}
\end{figure*}
Having shown that our model assumptions are adequately satisfied, we can now move on to testing the predictions from the model, based on the triad phase histograms collected. Figure \ref{fig:triads}(a) shows the PDF of two example triads, one with $\langle \CC \rangle <0$ (blue) and the other with $\langle \CC \rangle>0$ (orange), and the best fit of Eq. \eqref{eq:theta_PDF} shown as grey dashed lines. Notice the $y$-axis scale -- these PDFs are very close to being uniform, justifying our assumptions above that $|\langle \CC \rangle| \ll\DD$. Not only do we find excellent agreement with our model, whose PDF fit matches remarkably well, we also find that the position of the PDF's maximum matches our predictions based on the sign of $\langle \CC \rangle$. This, in turn, means that our prediction for the sign of the alignment $\langle \cos{\tphase}\rangle$ based on $\langle \CC \rangle$ holds. 
This doesn't just hold for the two triads shown, but also for every triad whose triad phase PDF is statistically converged.
We demonstrate this in the middle column of Figure \ref{fig:triads}, where we have taken the 200 most aligned triads and separated them based on the sign of $\langle \CC \rangle$ (shown in descending order of alignment magnitude).
 The maxima of each PDF agrees with what the theory predicts based on $\langle \CC \rangle$.
Equivalently, all statistically converged triads have a positive $\DD$ based on their PDF fits. We have therefore shown that our model for triad phase dynamics correctly captures the statistical distribution of triad phases and predicts the sign of the alignment, which will be an essential component of what we prove below about energy evolution and flux. An estimate for the magnitude of the alignment requires knowledge of $\DD$, which we found depends only on $\sqrt{p^2+q^2}$, in a roughly linear fashion (Fig. \ref{fig:triads}(d)).

\emph{Implications on energy evolution and flux---}Next we explore the implications of our model. Assuming that phase dynamics occur on faster time-scales than energy evolution \cite{Tennekes1975,LesieurBook}, we can approximate the dynamical evolution of $\rho_\kk$ (and therefore energy) using a quasi-static approach. After applying $\langle\cos(\tphase)\rangle \approx -\CC/(2\triad Dkpq)$ to the transfer function, Eq. \eqref{eq:transf_def}, we find
\begin{equation}
    \triad t k p q \approx (q^2-p^2)(\qq \times \pp)^2 \frac{\KK}{2 \triad Dkpq}, \label{eq:transf_closure}
\end{equation}
resulting in a closure of the energy equation in terms of $\rho_\kk$. $\KK$ doesn't change under cyclic permutation of triad vectors, and changes sign when exchanging any two vectors in a triad. The former property ensures energy conservation via $\triad tkpq + \triad tqkp +\triad tpqk = 0$, as well as enstrophy conservation via $k^2\triad tkpq + q^2\triad tqkp +p^2\triad tpqk = 0$. 
Although the exact magnitude of the transfer function depends on $\triad Dkpq$, the fact that $\triad Dkpq>0$ implies that the \textit{sign} of the transfer function is completely determined by $\KK$, a known quantity given the energy spectrum (Eq. \eqref{eq:Cdef}). 

We show now that the triad phase dynamics determine the sign of the average flux of conserved quantities in 2D turbulence, with no fitting parameters. 
Kraichnan's analytical work on 2D turbulence showed that energy and enstrophy cascade in opposite directions in their respective inertial ranges, namely $\text{sgn}(\Pi_E)=\text{sgn}(\triad t{\hat{k}}vw)$ and $\text{sgn}(\Pi_\Omega)=-\text{sgn}(\triad t{\hat{k}}vw)$, where $\hat{\kk},\mathbf{v}$ and $\mathbf{w}$ are members of a re-scaled triad such that $|\hat{\kk}|=1$ and $v<1<w$ \cite{Kraichnan1967}. Determining the sign of $\triad t{\hat{k}}vw$ therefore determines the sign of each flux.
Without a direct model for the transfer function, he argued that turbulence evolves towards its thermal equilibrium spectrum, and showed that a quasi-normal closure model gave the same sign.
Our statistical model of the dynamics of triad phases provides a direct prediction for the sign of $\triad t{\hat{k}}vw$ via Eq. \eqref{eq:transf_closure}. 
Since $w>v$, the sign of $\triad t{\hat{k}}vw$ is entirely determined by the sign of $\triad {\mathcal{K}} {\hat{k}}vw$, a property of the triad phase statistics. We assume an inertial range with an isotropic mean state and a corresponding one-dimensional energy spectrum satisfying $E(k) \propto k^{-n}$, where $E(k) = \sum_{|\kk|=k} |\kk|^2\rho_\kk^2 /2$. This implies $\rho_{\kk} \propto k^{-(n+3)/2}$. Plugging in this scaling into the definition of $\KK$ in Eq. \eqref{eq:Cdef} and dividing out $k$ we are left with:
\begin{equation}
    \triad {\mathcal{K}} {\hat{k}}vw \propto (w^2-v^2)-(1-v^2)w^{1+n}-(w^2-1)v^{1+n}. \label{eq:Kscaled}
\end{equation}
From this, it is straightforward to prove that $\triad {\mathcal{K}} {\hat{k}}vw < 0$ if $|n|>1$ (see End Matter). Therefore, we find that $\text{sgn}(\Pi_E)<0$ and $\text{sgn}(\Pi_\Omega) > 0$ in their respective inertial ranges of $n=5/3$ and $n=3$, as is observed in numerical and physical experiments (Fig. \ref{fig:spec_flux}(b)). 

The dynamics of the triad phase can also describe states in which no flux occurs. Such ``equilibrium'' states for 2D turbulence have an energy spectrum $E_{eq}(k) = 2 \pi k/(\gamma + \beta k)$, where $\gamma$ and $\beta$ are constants which determine the energy and enstrophy in the system, respectively \cite{Kraichnan1967,Kraichnan1975,AlexakisReview}.
Assuming $E(k)\propto E_{eq}(k)$ indeed results in $\KK=0$ and a triad phase equation which consists of only `random forcing' by neighbors. Our model therefore correctly captures the fact that an equilibrium state is described by random phases (equivalently, no alignment of triad phases).

\emph{Conclusions---}In this work, we have shown that the triad phase dynamics determine the presence, direction, and strength of cascades of conserved quantities in 2D turbulence. 
We introduced a simplified model for the triad phase dynamics based on the observation that triad phases are weakly correlated with their neighbors, and used a series of direct numerical simulations of 2D turbulence to show that our assumptions are justified and that the model correctly reproduces the statistics of triad phases. Our model retains the so-called self-interaction term in the triad phase dynamical equation, but treats the combined effect of neighboring triads as `noise'. In doing so, we significantly reduce the complexity of the evolution of the triad phase, allowing one to solve for its PDF, and therefore also for the expectation value of the cosine of the triad phase, which enters directly in the energy equation. Under a quasi-static assumption, our model produces a new closure for the energy evolution. At steady state, it determines the sign of the energy and enstrophy fluxes with no free parameters, and can be used to calculate the mean flux knowing the noise variance.

Our findings suggest that the sign of the triad phase alignment (and therefore flux) is determined by the linear stability properties of an \textit{isolated} triad phase. This is reminiscent of Waleffe's ``instability assumption'' \cite{Waleffe1992,Waleffe1993}, where flux direction was predicted solely from the linear stability properties of an isolated triad. While Waleffe's assumption lacked a clear \textit{a priori} justification\cite{Moffatt2014}, our analysis offers a plausible mechanism: sufficiently weak inter-triad phase correlations justify studying single-triad dynamics. Our work goes beyond the instability assumption by predicting not only the direction of the flux but also its expected magnitude via a novel closure for the energy equation. This closure could be seen as a `quasi-normal' closure \cite{Orszag1970,LesieurBook}, in that it closes the third moment of the complex velocities in terms of the second moment (the energy). However, it is not equivalent to any existing closure and does not rely on further assumptions such as the eddy-damped and Markovian properties typically applied to quasi-normal closures.

The predictive success of the model for triad phase dynamics relies on the assumption that neighboring triad phases are weakly correlated. We have found this to be true as long as no strongly coherent, energy-dominating vortices are present in the flow. These vortices can form in 2D turbulence when the forcing scale is not affected by dissipation (even in the case of a large-scale dissipation) and cause a strong deviation from the expected Kraichnan--Leith--Batchelor (KLB) similarity theory predictions of $n=5/3$ and $n=3$  \cite{Borue1994,Danilov2001,Scott2007,Burgess2017,Meunier2025}. The coherence of these vortices implies a strong correlation between different modes and therefore a breakdown of the expected triad phase PDFs.  We leave investigations of this breakdown and its impact on the inertial ranges of 2D turbulence to future work.

Overall, we have shown how the dynamics of the triad phases contain essential information about the cascades of conserved quantities in 2D turbulence. Triad phases are not unique to 2D turbulence, and indeed previous studies have shown the important role that triad phases play in the transfer and cascades of conserved quantities in other systems. It is therefore possible that simple descriptions of triad phase dynamics in these other contexts can be found.
Beyond fluid models, our approach could be useful to describe the mean behavior of other nonlinear PDEs displaying non-equilibrium regimes, just as the wave turbulence approach has been applied to a diverse set of systems.

\emph{Acknowledgments---}
The authors would like to thank Luca Biferale, Michele Buzzicotti, Anna Frishman, \"{O}zg\"{u}r G\"{u}rcan, Javier Jim{\'e}nez and Adrian van Kan for helpful discussions during the development of this project and manuscript.
S.J.B. has received funding from HORIZON EUROPE Marie Sk{\l}odowska-Curie Actions (H2022, Grant No. 101109237).

\emph{Code Availability---}
The code used to run the simulations of two-dimensional turbulence described in the main text is v1.1 of `\verb|2DHD_GPU|' \cite{Benavides2026CodeGPU}. It is preserved at \url{https://doi.org/10.5281/zenodo.20024314}, available via the Creative Commons Attribution 4.0 International (CC-BY) license.

\emph{Data Availability---}
The Figshare repository \cite{Benavides2026Data}, preserved at \url{https://doi.org/10.6084/m9.figshare.32159295}, contains (i) the full run directory, including the dataset, from the ensemble simulation of 2D turbulence described in the text; (ii) post-processed temporal data used in Fig. \ref{fig:noise}, including the empty run directories for reproducing the time series; (iii) the scripts used to generate the figures in the main text.


\appendix
\section*{End Matter}
\emph{Simulation details---} Simulations of two-dimensional turbulence in a periodic domain were performed using a standard pseudo-spectral method with $2/3$ dealiasing and a Runge-Kutta method (2nd order) with an adaptive time step for time integration. The code is written in Python, and uses CuPy (version 13.4) to provide GPU acceleration \cite{CuPy}. It is capable of running independent ensemble members (handling all members in one 3D array) and collecting triad phase and amplitude statistics. The code used to run the 2D turbulence simulations is available on GitHub at \cite{Benavides2026CodeGPU}. The foundation of the pseudo-spectral code is based on a parallel CPU Fortran code written by Pablo Mininni (Universidad de Buenos Aires) \cite{Mininni2011}. 

We simulated the streamfunction-vorticity formulation of the Navier Stokes equations in a square domain with side lengths of $2 \pi$ using $512^2$ grid points. We included regular viscosity $\nu \nabla^2 \vv$, with $\nu = 0.0015$, and a large-scale drag $-\mu \vv$, with $\mu=0.1$. We forced the streamfunction with a constant-in-time function $f(\mathbf{x})$ with a mean `energy' $\sqrt{\int|k^2 f(\mathbf{x})|^2 \ d^2\mathbf{x}}=10.6$. The forcing function is composed of spectral modes satisfying $k_f - 0.5 < k < k_f + 0.5$, where $k_f = 24$, each of which is multiplied by a random phase. The parameter values chosen result in an averaged Reynolds number $Re = v_{k_f,\mathrm{rms}}/(k_f \nu) = 13.8$ and a drag-based Reynolds number $Re_\mu = k_fv_{k_f,\mathrm{rms}} / \mu = 119.1$, where $v_{k_f,\mathrm{rms}}$ is the square root of the time- and space-mean of the energy based on a velocity spectrally filtered around $k_f -0.5 < k < k_f + 0.5$. Runs were initialized with modes $k<k_f$ and random phases.

We performed an ensemble of ten independent simulations of 2D turbulence. Each ensemble member was provided a different random seed, resulting in different initial conditions and forcing functions for each ensemble member. Statistics, such as spectra, flux, and triad quantities, were gathered once the run reached steady state. 
We collected time-series data of $\tphase$ and the hypothetical `noise' variable $\XXs := \sum_{\mathbf{k^\prime} = \{\kk,\pp,\qq\}}\sum^\Delta_{\mathbf{l},\mathbf{r}} \triad {N}{k^\prime}lr \sin{(\triad \theta {k^\prime} lr)}$ for 48 triads, and computed on-the-fly histograms of $\tphase$ for 8657 triads. The triads considered were all randomly chosen, and were taken from both energy and enstrophy cascade ranges, but did not have modes in both ranges. The script used to create the list of triads can be found in the GitHub repository of the simulation code \cite{Benavides2026CodeGPU}. 
Fluctuations dominate the triad phase dynamics, requiring very long integration times for converged statistics ($\approx 4 \times 10^9$ time steps or $\approx 10^7$ turnover times per ensemble member). This has restricted our resolution to $512^2$ (with $2/3$ dealiasing), resulting in relatively weak inertial ranges with non-constant fluxes of conserved quantities. However, our model for triad phase statistics does not require the presence of an inertial range (indeed, dissipation does not enter into the phase dynamics), and we focus on the direction of the cascades.

\emph{L\'{e}vy noise case---} 
We consider $\XX$ in Eq. \ref{eq:noisyAdler} to be a white L\'{e}vy noise with an $\alpha$-stable distribution with $\alpha=1.8$ and zero skewness. The generalized Fokker-Planck equation associated with the Langevin equation is \cite{ChechkinBook}
\begin{equation}
    \frac{\partial \mathcal{P}}{\partial t} = -\frac{\partial}{\partial \tphase}\left(\CC\sin(\tphase)\mathcal{P}\right) + \triad {\mathcal{D}}kpq \frac{\partial^\alpha \mathcal{P}}{\partial |\tphase|^\alpha}, \label{eq:levy_FP}
\end{equation}
where the last term on the right-hand-side is the symmetric Riesz space fractional derivative \cite{Saichev1997}, most easily defined by its Fourier transform $\mathcal{F}$:
\begin{equation}
    \mathcal{F}\left\{\frac{\partial^\alpha \mathcal{P}}{\partial |\tphase|^\alpha} \right\} = - |m|^\alpha\widehat{\mathcal{P}}(m),
\end{equation}
where $m$ is the mode number (since $\tphase$ is $2\pi$-periodic), and $\widehat{\mathcal{P}}=\mathcal{F}\{\mathcal{P}\}$ is the Fourier transform of the PDF. Taking the Fourier transform of Eq. \eqref{eq:levy_FP} we find,
\begin{align}
    \frac{\partial \widehat{\mathcal{P}}(m)}{\partial t} = -\frac{m\CC}{2}\left(\widehat{\mathcal{P}}(m-1) -\widehat{\mathcal{P}}(m+1)\right)& \nonumber\\ - \triad {\mathcal{D}}kpq|m|^\alpha\widehat{\mathcal{P}}(m),&
\end{align}
with the condition that $\widehat{\mathcal{P}}(0) = 1/2\pi$ and $\widehat{\mathcal{P}}(-m)=\widehat{\mathcal{P}}^*(m)$, where $*$ denotes the complex conjugate. To find the steady state PDF we set the left-hand side to zero. Without any further assumptions, this gives a recursive relation for the solution $\widehat{\mathcal{P}}(m)$. Instead of pursuing a closed-form solution for this, we will make the assumption that $|\CC| \ll \triad {\mathcal{D}}kpq$, based on observations from simulations. Treating $\varepsilon := |\CC|/\triad {\mathcal{D}}kpq \ll 1$ as the small parameter and expanding our solution using an asymptotic series ansatz, $\widehat{\mathcal{P}}(m) = p_0(m)+\varepsilon p_1(m)+\cdots$, we can then balance at different orders of $\varepsilon$. At zeroth order, we find $0=-|m|^\alpha p_0(m)$, which tells us that $p_0(0) = 1/2\pi$ and $p_0(m>0)=0$ (a uniform distribution). At next order, the equation becomes $m(p_0(m-1)-p_0(m+1))=-2|m|^{\alpha}p_1(m)$. The only nonzero $p_0(m)$ is $p_0(0)$, so that $p_0(0) = -2 p_1(1)$ and $p_0(0) = -2p_1(-1)$ are the only two nontrivial equations at this order. Thus, $p_1(1) = p_1(-1) = -1/4\pi$ and the Fourier representation of $\mathcal{P}$ becomes $\mathcal{P}(\tphase)=1/2\pi - \varepsilon e^{i\tphase}/4\pi - \varepsilon e^{-i\tphase}/4\pi $. Up to first order, the resulting steady state PDF is
\begin{equation}
    \mathcal{P}(\tphase) \approx \frac{1}{2\pi}\left(1- \frac{\CC}{\triad {\mathcal{D}}kpq}\cos\left(\tphase\right)\right). \label{eq:levy_PDF_approx}
\end{equation}
Consider now the PDF found assuming $\XX$ is a white Gaussian noise with variance $\triad Dkpq$ (Eq. \eqref{eq:theta_PDF}). Taylor expanding around $\CC/\triad Dkpq = 0$ and noting that $I_0[\CC/\triad Dkpq]\rightarrow1$ as $\CC/\triad Dkpq\rightarrow 0$,  we see that, up to first order, the two PDFs \eqref{eq:theta_PDF}  and \eqref{eq:levy_PDF_approx} are the same, but with $\triad {\mathcal{D}}kpq$ replaced with $\triad Dkpq$. We therefore reach the conclusion that the effect of L\'{e}vy noise, which produces fractional diffusion of the PDF, only enters at second order in $\CC/\triad {\mathcal{D}}kpq$. Given that $|\CC|$ is much smaller than the measured noise variance in our simulations of 2D turbulence, we choose to consider the white Gaussian noise case for our model for the sake of simplicity.

\emph{Sign of $\triad {\mathcal{K}}{\hat{\kk}}vw$---} In this section, we prove that $\triad {\mathcal{K}}{\hat{\kk}}vw < 0$ if $|n|>1$. It is similar in approach to a proof done by Kraichnan for a different quantity associated with a rescaled triad \cite{Kraichnan1967}. We begin with $\KK$ in Eq. \eqref{eq:Cdef}. As stated in the main body, we assume that $E(k) \propto k^{-n}$, so that $\rho_\kk = \rho_0 k^{-(n+3)/2}$. If we plug in the functional form of $\rho_\kk$ into the definition of $\KK$, we get:
\begin{eqnarray*}
    \frac{\KK}{\rho_0^4} &=& \frac{q^2-p^2}{k^2}(p q)^{-(n+3)} +  \frac{p^2-k^2}{q^2}(k p)^{-(n+3)} \\ 
    &&+  \frac{k^2-q^2}{p^2}(k q)^{-(n+3)}.
\end{eqnarray*}
If we consider $p<k<q$, we define the rescaled triad so that $p=kv, q = wk$, with $v<1<w$. After making these substitutions, we find:
\begin{eqnarray}
    \frac{\KK}{k^{-(n+3)}\rho_0^4} &=& (w^2-v^2)(vw)^{-(n+3)} +  \frac{v^2-1}{w^2}v^{-(n+3)} \nonumber\\  
    &&+  \frac{1-w^2}{v^2}w^{-(n+3)}. \label{eq:app_Kscaled}
\end{eqnarray}
Dividing the right-hand-side of Eq. \eqref{eq:app_Kscaled} by $(vw)^{-(n+3)}>0$ results in a slightly simpler version which we will call $F(v,w,n)$ for the remainder of this section, namely: 
\begin{equation}
    F(v,w,n) = (w^2-v^2) +  (v^2-1)w^{1+n} +  (1-w^2)v^{1+n}. \label{eq:app_Fdef}
\end{equation}
This gets us to Eq. \eqref{eq:Kscaled}, the RHS of which is the quantity whose sign we are trying to determine. Since $\text{sgn}(\triad {\mathcal{K}}{\hat{\kk}}vw) = \text{sgn}(F(v,w,n))$, our goal is to show that $F(v,w,n(n))<0$ for $|n|>1$.

We begin by noting its zeros, $F(1,w,n) = F(v,1,n) = 0$, which tells us where the function changes sign. To determine the sign of $F$ in the region of interest, $v<1<w$, we look at $\partial F/\partial v$. We will show that the sign of $\partial F / \partial v$ is fixed for all $v$ and $w>1$, but depends on $n$, thereby determining the sign of $F(v,w,n)$. Now,
\begin{equation}
    \frac{\partial F}{\partial v} = v\left[2\left(w^{1+n}-1\right) - (1+n)\left(w^2-1\right) v^{n-1}\right], \label{eq:app_Fprime}
\end{equation}
whose sign only depends on the term inside the square brackets, but is not immediately apparent. Consider $G(w) = 2(w^{1+n}-1) - (1+n)(w^2-1)$. Since $G(1) = 0$, to get the sign of $G(w)$ for $w>1$, we look at $\partial G / \partial w = 2(1+n) (w^n-w)$. Given that $w>1$, the term containing $w$ is positive if $n>1$ and negative otherwise, whereas the $(1+n)$ coefficient changes sign when $n<-1$, giving us the result that $\partial G / \partial w > 0$ if $|n|>1$. Since $\partial G / \partial w > 0$ and $G(1) = 0$, we then know that $G(w) > 0$ for $|n|>1$ and $w>1$. The fact that $G(w)>0$ implies that $2(w^{1+n}-1) > (1+n)(w^2-1)$. We can use this result to give a bound on Eq. \eqref{eq:app_Fprime}:
\begin{eqnarray*}
    \frac{\partial F}{\partial v} &=& v\left[2\left(w^{1+n}-1\right) - (1+n)\left(w^2-1\right) v^{n-1}\right], \\
    &>& v(1+n)\left(w^2-1\right)(1 - v^{n-1}), \quad \text{if} \quad |n|>1, \\
    &>& 0, \quad \text{if} \quad |n|>1, \\
\end{eqnarray*}
where the last inequality results from the observation that, if $n>1$, $(1+n)$ and $(1-v^{n-1})$ are both positive, whereas if $n<-1$ both will be negative. We can then finally conclude that, since $F(1,w,n) = 0$ and $\partial F / \partial v>0$, then $F(v,w,n) < 0$ for $v<1<w$ and $|n|>1$. This tells us that $\triad {\mathcal{K}}{\hat{\kk}}vw < 0$ for $|n|>1$, and $\triad {\mathcal{K}}{\hat{\kk}}vw > 0$ for $|n| < 1$. Note also that $F(v,w,\pm1)=0$ so that $\triad {\mathcal{K}}{\hat{\kk}}vw = 0$ for $n = \pm1$ (this corresponds to the result on equilibrium ranges).

\bibliography{phases_2D}

\end{document}